# Highlights

## Encouraging Emotion Regulation in Social Media Conversations through Self-Reflection

Akriti Verma,Shama Islam,Valeh Moghaddam,Adnan Anwar

- Introduces a model for on-the-spot attention and response modulation support in online conversations by encouraging self-reflection in moments of ongoing highly elevated emotional expression.
- Proposes a graph-based framework for identifying the need for emotion regulation in online social media conversations.
- Presents design implications for social media applications to incorporate support for users' emotion regulation.

# Encouraging Emotion Regulation in Social Media Conversations through Self-Reflection★,★★

Akriti Verma[a,∗], Shama Islam[a], Valeh Moghaddam[b] and Adnan Anwar[b]

[a]*School of Engineering, Deakin University, Geelong, 3216, Australia*
[b]*School of Information Technology, Deakin University, Geelong, 3216, Australia*

**ABSTRACT**

Anonymity in social media platforms keeps users hidden behind a keyboard. This absolves users of responsibility, allowing them to engage in online rage, hate speech, and other text-based toxicity that harms online well-being. Recent research in the field of Digital Emotion Regulation (DER) has revealed that indulgence in online toxicity can be a result of ineffective emotional regulation (ER). This, we believe, can be reduced by educating users about the consequences of their actions. Prior DER research has primarily focused on exploring digital emotion regulation practises, identifying emotion regulation using multimodal sensors, and encouraging users to act responsibly in online conversations. While these studies provide valuable insights into how users consciously utilise digital media for emotion regulation, they do not capture the contextual dynamics of emotion regulation online. Through interaction design, this work provides an intervention for the delivery of ER support. It introduces a novel technique for identifying the need for emotional regulation in online conversations and delivering information to users in a way that integrates didactic learning into their daily life. By fostering self-reflection in periods of intensified emotional expression, we present a graph-based framework for on-the-spot emotion regulation support in online conversations. Our findings suggest that using this model in a conversation can help identify its influential threads/nodes to locate where toxicity is concentrated and help reduce it by up to 12%. This is the first study in the field of DER that focuses on learning transfer by inducing self-reflection and implicit emotion regulation.

## 1. Introduction

The practice of consciously modifying one's affective state is called emotion regulation. The ability to successfully perform emotion regulation is essential to function effectively in everyday life, to act appropriately in everyday interactions, or merely for hedonic purposes Wadley, Smith, Koval and Gross (2020). The topic has been thoroughly explored in the field of psychological work, the study of cognitive behaviour as well as mental health McRae and Gross (2020), Gross (2015), Gross (2014). Owing to the boost in technology and access to digital media which provides a wide range of options available at ease, this practice of regulating emotions through the use of digital media has seen tremendous growth recently.

Digital technologies provide a greater range of strategic options that can be easily and effectively executed. Individuals combine a variety of applications and devices for purposefully managing emotions in daily life Smith, Wadley, Webber, Tag, Kostakos, Koval and Gross (2022). Some examples include listening to uplifting music while exercising, watching comedy or light-hearted videos to relieve stress after work, playing social video games when feeling lonely or scrolling through social media applications to combat boredom. Social media applications are widely used by people, multiple times throughout the day. These applications contain several emotional affordances (expressible, shareable, consumable, and assessable), all of which can influence emotions as well as a behaviour associated with emotions Steinert and Dennis (2022). Owing to its vast usage, activities on social media applications significantly impact online well-being. The prevalence of toxicity and hate speech in online conversations has been largely observed and studied in recent years and has been found to be a crucial element of virality Maarouf, Pröllochs and Feuerriegel (2022), Goel, Anderson, Hofman and Watts (2016) which is a measure of a post's reach. Social media conversations are fuelled by connective action and fast information spread and have given rise to online movements and debates, the results of which have affected offline events Saveski, Roy and Roy (2021), Mirbabaie, Brünker, Wischnewski and Meinert (2021). Recently, people have started being vocal about how the hate received online impacts their daily lives and questions their safety online. Posts from political people in power, news websites and young content creators, to name a few, are victims of this. It has been discovered that encountering or dealing with disrespectful or rude behaviour online is now considered standard and a part of the deal Thomas, Kelley, Consolvo, Samermit and Bursztein (2022). There have been some rules enforced by social media applications where accounts with a large number of followers and engagement were banned to curb the spread of offensive speech, but toxicity is still prevailing as it arises from the actions of many mildly toxic people as opposed to a few highly toxic ones Saveski et al. (2021).

★
★★

∗Corresponding author

vermaakr@deakin.edu.au (A. Verma); shama.i@deakin.edu.au (S. Islam); valeh.moghaddam@deakin.edu.au (V. Moghaddam); adnan.anwar@deakin.edu.au (A. Anwar)

ORCID(s): 0000-0003-3963-0870 (A. Verma)

It is essential to ensure that social media platforms offer a safe space for healthy interaction and communication by managing the vulnerabilities to digital wellbeing. Previous studies in the field of social media and emotion regulation have focused on how social media facilitates online emotions and their effects on eudaimonic well-being Goldenberg and Gross (2020), Steinert (2021), Yang, Liu, Li and Shu (2020), Yue, Zhang and Xiao (2022), Shi, Koval, Kostakos, Goncalves and Wadley (2023). These include investigating how maladaptive emotion regulation methods affect problematic use of social media and smartphones, as well as determining how informed smartphone use can improve effective emotion regulation and social competence Zsido, Arato, Lang, Labadi, Stecina and Bandi (2021). Prior research has also found that depending on the automaticity and situational social media aspects, both active and passive social media use could be procrastination or recovery activities Hossain, Wadley, Berthouze and Cox (2022). Ments, Treur, Klein and Roelofsma (2021) present a second-order adaptive brain network model for simulating the process of emotion regulation in social media and discover how, while some emotion regulation strategies are protective in the short term, using them consistently results in worsened mood and relatively low well-being. Studies have also examined the user interfaces of social media applications and recommended design frameworks to assist emotion regulation in breaking the habit of making unpleasant comments on social media platforms by automatically identifying emotional aspects, such as the audience's anticipated emotional response to users' comments Kiskola, Olsson, Väätäjä, H. Syrjämäki, Rantasila, Isokoski, Ilves and Surakka (2021).

Although these developments provide significant insights into how the process of emotion regulation unfolds in social media applications, there is a lack of digital solutions available for implementation. Current tools for emotion regulation include mood-based recommendation systems and reminders, which can only provide temporary assistance and are difficult to incorporate into daily life Wadley et al. (2020), Slovak, Antle, Theofanopoulou, Roquet, Gross and Isbister (2022). Additionally, there is a lack of a common prototype for synthesising emotion regulation because the majority of recent research in the field of digital emotion regulation is based on field studies, ecological momentary assessments (EMA), or physiological sensors combined with facial data, making it difficult to extend and replicate for further research Ruensuk, Cheon, Hong and Oakley (2020). It is also necessary to understand how to identify the need for emotion regulation in online environments. This entails the creation of a solution that informs users of their micro-impacts on a post, rather than providing a broad overview of "what may be the consequence." This information about one's impact will call their anonymity on a post into question and encourage them to act responsibly.

This work presents an innovative approach to delivering information regarding the need to regulate one's emotions and guiding them through the emotion regulation experience, in social media conversations, intending to embed the learning into their lives through repetitive application and, as a result, enhance online well-being. Because this study was conducted using publicly available data, it provides a foundation for expansion, extension, comparison, and contrast. Therefore, the main contributions of this work are:

- We introduce a model for on-the-spot attention and response modulation support in online conversations by encouraging self-reflection in moments of ongoing highly elevated emotional expression.

- We propose a graph-based framework for identifying the need for emotion regulation in online social media conversations.

- We present design implications for social media applications to incorporate support for users' emotion regulation.

The remainder of this paper is organised as follows. We begin by discussing recent advances in the literature, followed by a description of the problem statement and proposed methodology. Then, we provide a detailed account of our experimental analysis and performance evaluation before concluding.

## 2. Related work

### 2.1. Digital Emotion Regulation

Recent DER research falls into three categories: observational studies, novel DER tools, and ER recognition using multi-modal sensors. Observational studies examine how people use social media apps such as video streaming platforms, discussion forums, online games, and music to regulate their emotions, thoughts, and behaviours Smith et al. (2022), Shen and Cox (2020). Several diary studies have been undertaken to better understand various elements of everyday emotion regulation in isolation, such as the usage of social media to overcome homesickness and university students' use of music streaming platforms Wadley, Krause, Liang, Wang and Leong (2019). Individuals' multitasking and passive scrolling habits on social media apps have also been investigated in studies revealing how people voluntarily take breaks from social media to mitigate negativity or maintain a sense of equilibrium, as well as highlighting the practice of interpersonal emotion regulation in discussion forums Hossain et al. (2022), Lukoff, Yu, Kientz and Hiniker (2018). The majority of studies on how digital media is used to regulate emotions have relied on self-reported questionnaires or a diary-keeping technique Smith et al. (2022), Shen and Cox (2020), Tag, van Berkel, Vargo, Sarsenbayeva, Colasante, Wadley, Webber, Smith, Koval, Hollenstein et al. (2022a), Wadley et al. (2019). Participants in this kind of data collection are required to document their interactions with emotion regulation and technology use over a given period, and then discuss their use of technology and the emotional reactions that accompany it in an interview. By limiting the amount of data that may be recorded, this strategy enables participants to track and reflect on significant insights from their work and social lives. Overall, these studies reveal that people use a variety of digital technologies for emotion regulation in everyday life, emphasising the importance of the technology packed inside these devices and the need to promote well-being online Wadley et al. (2020).

**Table 1**
Digital interventions for ER, classified according to the strategy they support

| ER Strategy Family | Novel Tools |
|---|---|
| Situation Selection: avoiding a situation that is likely to provoke unpleasant emotions. | <ul><li>Augmented Reality-based ER technologies to portray affect as interaction states to facilitate interpersonal ER Semertzidis, Scary, Andres, Dwivedi, Kulwe, Zambetta and Mueller (2020).</li><li>Critical Voice in User Interface Design for ER in Online Conversations Kiskola et al. (2021), Kou and Gui (2020).</li></ul> |
| Situation Modification: modifying a scenario's characteristics to alter its emotional impact. | <ul><li>Robot-based artificial commensal companions to facilitate social interactions while eating for people who want or are forced to eat alone Mancini, Niewiadomski, Huisman, Bruijnes and Gallagher (2020). Interaction-based technology to enable users up-regulate positive emotions Li, Hao and Yoon (2020)</li><li>Soundscapes for enhancing task performance and mood Newbold, Luton, Cox and Gould (2017), Yu, Hu, Funk and Feijs (2018).</li></ul> |
| Attentional Deployment: focusing on or away from emotional elements in a situation to evoke the intended emotion. | <ul><li>Customisable virtual reality environments for in-the-moment soothing support for open workplaces Ruvimova, Kim, Fritz, Hancock and Shepherd (2020).</li><li>Photos with expression and intent to achieve calm/boost desired emotions Chen, Mark and Ali (2016).</li></ul> |
| Cognitive Change: reevaluating a scenario to reshape its emotional effect. | <ul><li>Personalised breathing pacer to reduce anxiety by inducing explicit ER, primarily used for distraction/reappraisal Miri, Jusuf, Uusberg, Margarit, Flory, Isbister, Marzullo and Gross (2020).</li><li>Haptics-based smartwatch intervention for cognitive, physiological, and behavioural changes Costa, Guimbretière, Jung and Choudhury (2019).</li></ul> |
| Response Modulation: transforming a current emotional reaction or expression into a more desired one. | <ul><li>Innovative toys to help school students improve their ER practices and implicit emotional beliefs through repeated interaction Theofanopoulou, Isbister, Edbrooke-Childs, Slovák et al. (2019).</li><li>Haptics-based guided breathing and pleasant scents to encourage safer driving Paredes, Zhou, Hamdan, Balters, Murnane, Ju and Landay (2018), Dmitrenko, Maggioni, Brianza, Holthausen, Walker and Obrist (2020).</li></ul> |

The second category includes the design and development of new tools for DER Kiskola et al. (2021), Kou and Gui (2020), Smith et al. (2022). This field of research focuses on the creation of interventions that aim to support, improve, or guide emotion regulation skills, or to assist individuals in applying such skills in challenging situations. The process of emotion regulation occurs in four stages: recognising the need or realising the desire for emotion regulation, selecting an appropriate strategy, applying it, and then monitoring the regulated state to determine whether additional regulation is required Wadley et al. (2020). Technology-enabled interventions have either intended to assist a specific ER strategy or to increase emotional awareness during the identification or monitoring stages Smith et al. (2022), Slovak et al. (2022). Table-1 lists recent interventions ER based on the strategy they support. They comprise experience-based design elements that largely rely on bio-feedback or implicit target responses to guide users subconsciously into specific physiological states via haptic contacts, such as simulating heart rate to improve performance by reducing anxiety Miri et al. (2020), Newbold et al. (2017), Mancini et al. (2020), Paredes et al. (2018). Recent advances have also seen the development of didactic intervention components that rely on reminder-based recommender systems, such as recommending specific ER techniques to users and encouraging emotion awareness by prompting users to examine how they feel or felt Chen et al. (2016), Costa et al. (2019). These works explore new design opportunities for DER and give a fresh set of directions for this field of study by investigating how these designs affect users.

**Table 2**
A summary of recent studies aimed at recognising Emotion Regulation in online environments

| Title | Purpose | Data used/ parameters monitored | Availability of the dataset | Size of the dataset | Availability of a reproducible model |
|---|---|---|---|---|---|
| Emotion trajectories in smartphone use: Towards recognizing emotion regulation in the wild  Tag, Sarsenbayeva, Cox, Wadley, Goncalves and Kostakos (2022b) | Present findings from a field study that measured how joy unfolds during everyday smartphone use. | Collected using a customised smartphone application that tracked physical (facial) features | Not available | The study involved 20 participants, was carried out for 14 days | NA |
| Benchmarking commercial emotion detection systems using realistic distortions of facial image datasets  Yang, Wang, Sarsenbayeva, Tag, Dingler, Wadley and Goncalves (2021) | Evaluate the performance of commercial emotion detection services | Utilised 3 facial expression based datasets | Used online datasets (ADFES) (RaFD), and (WSEFEP) | 838 pictures in total | NA |
| Behavioural and Physiological Signals-Based Deep Multimodal Approach for Mobile Emotion Recognition  Luo and Yang (2021) | Propose a novel attention-based LSTM system that uses a combination of sensors (front camera, microphone, touch panel) from a smartphone and wristband | Collected a dataset using a smartphone application where the behavioural and physiological parameters were taken into observation | Not available | The study involved 45 participants | Described in the paper |
| How Do You Feel Online? Exploiting Smartphone Sensors to Detect Transitory Emotions during Social Media Use  Ruensuk et al. (2020) | Explore the identification of people's emotions when they use social media applications | Collected a dataset using various (physical) motion /eye-tracking applications | Not available | The study involved 20 participants | NA |
| **Encouraging Emotion Regulation in Social Media Conversations through Self-Reflection** | **Identify the contexts that need ER and provide on the spot support for the same** | **Utilised data from Twitter conversations** | **Publicly available** | **Size of the dataset: 180K** | **Described in the paper** |

The third category of studies includes interventions to recognise and capture the process of DER using multi-model sensors. Apart from an individual's desire to regulate their emotions, the ER process includes their surroundings, a situational trigger for emotion, and their attempts to regulate that emotion. Because measuring these features in a lab setting is challenging, research has begun to look into ways to recognise them in the wild. The front camera of smartphones, touch sensors, eye trackers, and motion sensors was utilised independently and in combination to observe the change of emotions using facial expressions Luo and Yang (2021), Ruensuk et al. (2020). A recent study used the device's front camera to measure people's levels of delight throughout each phone session and put them into three groups based on how likely they were to feel joy at the start of the session Tag et al. (2022b). Another study employed image manipulation to imitate the actual image distortion that occurs when capturing a person's expressions for facial expression-based detection and regulation of digital emotions Yang et al. (2021). Combining modalities, according to studies, improves the accuracy of affect detection in social media tasks. Table-2 summarises recent studies aimed at identifying emotion regulation using sensors and digital media. As can be seen, the majority of these studies rely on data collected specifically for the study, making it difficult to expand on. As a result, this paper presents a framework for identifying and supporting ER using publicly available data.

### 2.2. Analysing Social Media Conversations

The growing popularity of social media applications has resulted in considerable virtualization of our life's engagement activities. Traditional means of leisure and entertainment have also been revolutionised by this digitalisation, as these applications enable a range of dimensions for expression and consumption. Social media offers a platform to facilitate robust online interactions; unfortunately, their potential is frequently hampered by the toxicity induced by atrocious speech and antisocial behaviour. Toxicity in social media platforms has been extensively studied in a number of existing research papers. Saveski et al investigated the individual and group structures of toxic conversations and discovered that, while toxicity is distributed among a large number of low to moderately toxic users at the individual level, toxic conversations at the group level have larger, wider, and deeper response trees but scattered follow graphs. They also notice that users with no social connection to the poster and few mutual friends are more likely to post toxic responses Saveski et al. (2021). Solovev et al examine the dynamics of virality, misinformation, and rumour propagation on social media, discovering that moral emotion encoded in source tweets influences the dissemination of false rumours on social media Solovev and Pröllochs (2022). Mirbabaie et al study the significance of social media during online social movements within the context of connective action theory by categorising influential users into roles and identifying connective action starters and maintainers Mirbabaie et al. (2021). These findings help us understand the social factors that contribute to toxic online behaviour and help to create healthier social media platforms. They also propose exposing the social context of the conversation to perhaps encourage respectful behaviour online.

### 2.3. Analysing Emotions in Social Media Conversation Graphs

Emotions in text-based social media interactions have been intensively studied using graphs, in which each node represents a user or a comment made by a user, and the edges reflect the link between users or comments. This procedure usually consists of two parts. The first is the emotional classification of text within a node, while the second is a node's structure-based connectivity and influence. Several Natural Language Processing (NLP)-based algorithms for recognising the emotion associated with a text have been proposed in the literature Poria, Majumder, Mihalcea and Hovy (2019), Chowanda, Sutoyo, Tanachutiwat et al. (2021), Majumder, Poria, Hazarika, Mihalcea, Gelbukh and Cambria (2019). These models have been presented for recognising the 6 primary emotions (love, joy, sadness, surprise, anger and fear) as well as the 27 secondary emotions Shaver, Schwartz, Kirson and O'connor (1987). The NRC lexicon is a crowdsourced library containing 27,000 English words and their associations with eight basic emotions (anger, fear, anticipation, trust, surprise, sadness, joy, and disgust) and two sentiments (negative and positive), is recently being used as a benchmark to measure emotional affect from a body of text Mohammad and Turney (2013).

The emotion classification result is fed into the graph and used as a node attribute, either for statistical analysis (such as centrality measures, distance from the root node, and the number of occurrences) or in the form of a Graph Neural Network (GNN). Individual users' influence, popularity, and social impact are frequently measured using statistical attributes-based analysis, but the neural network-based category analyses the social graph as a whole in order to understand its structure and high-level network dynamics Antonakaki, Fragopoulou and Ioannidis (2021). In their proposed implicit sentiment analysis model, Yang et al. used a graph attention convolutional neural network to propagate semantic information and an attention mechanism to calculate the contribution of words to emotional expression Yang, Xing, Li and Chang (2022). Perikos et al devised a graph-based method for modelling a topic's emotional level based on emotions, a topic's emotion graph is then created to show the public's feelings and mood regarding the subject Perikos and Hatzilygeroudis (2018). Brambilla et al offered a novel method for retrieving common patterns in online conversations by employing a directed multigraph network to develop and understand communication patterns among users, beginning with the hierarchical structure of posts and comments Brambilla, Javadian and Sulistiawati (2021). These studies demonstrate that by modelling social media conversations in the form of graphs, their structural, as well as dynamic features, can be explored.

### 2.4. Identifying the Context when Emotion Regulation is Needed

Disrespectful commentary contributes significantly to online toxicity. Content moderation has recently been identified as an intervention to enhance online well-being and minimise toxicity by recognising uncivil remarks based on keywords or underlying emotions Thomas et al. (2022), Jhaver, Boylston, Yang and Bruckman (2021). Scalable implementations of machine learning-based moderation approaches have also been explored Gorwa, Binns and Katzenbach (2020), Gillespie (2020). Algorithmic solutions can be utilised to display users a content analysis (emotion-based) of published comments, which may cause some users to reconsider their posts. For example, the Perspective API Google (2021), which detects toxic writing as a percentage score from a body of text, when implemented into the comment writing system of the Spanish language news site El Pas, was found to have moderately enhanced the quality of discussions. Internet etiquette and "free expression" guidelines can be ambiguous., which is why the goal here is to inform the user of the implications of their comments. This will improve their ability to empathise with other online users and trigger implicit emotion management Walther (1993). We argue that the challenge of effective ER in computer-mediated textual communication can be met by presenting factual cues to users, as suggested by Kiskola et al, i.e. supporting emotion regulation through automatic identification of emotional elements, and Slovak et al, who highlight the need for DER interventions that would integrate didactic learning into the lives of the participants Kiskola et al. (2021), Slovak et al. (2022). The topic of implicit emotion regulation has recently received attention in the scientific literature. In contrast to explicit emotion regulation, which involves consciously suppressing emotional reactions, implicit emotion regulation is automatic and may be automated Torre and Lieberman (2018). As a result, within the framework of this research, implicit emotion regulation appears as a potential design approach. Emotion regulation can be enhanced by affect labelling, which simply makes emotionally charged aspects of a conversation more apparent. Kiskola et al. provide critical viewpoints on potential solutions by presenting and discussing systems that are meant to promote emotion regulation through self-reflection Kiskola et al. (2021). In this work, we propose to notify the user of their emotional impact on the conversation. Therefore, by providing "what has happened" as a result of a comment, this work suggests an alternative to discarding "what may happen" based on a comment.

## 3. Problem Statement and Proposed Methodology

### 3.1. Problem Statement

Numerous platforms, user roles, how engagement develops, and what sustains them have all been widely studied in relation to hate speech in online conversations Solovev and Pröllochs (2022), Saveski et al. (2021), Majó-Vázquez, Nielsen, Verdú, Rao, de Domenico and Papaspiliopoulos (2020), Guberman, Schmitz and Hemphill (2016), Konikoff (2021). Recently, it has become popular to recognise the presence of emotion regulation as a component of hate speech. In this work, we aim to target the hate speech generated by digital emotion dysregulation in online conversations. Since emotion regulation is a subjective process that depends on context, which cannot always be inferred online, we propose that by analysing people's online actions, we can identify the contexts where emotion regulation is needed and subsequently support them in the process by providing information that encourages self-reflection. Even though we cannot be certain that a user is actively attempting to regulate their emotions at any given moment, we believe implicit emotion regulation, aided by design cues and labelled factual analysis (rather than predictions), will enable users to more easily understand the effects of their actions and consider posting thoughtful comments. Therefore, in this research we try to answer the following research questions:

- How can we determine when online conversations require emotion regulation?
- How can we provide on-spot support for emotion regulation in online conversations, extending beyond pattern detection?
- How can we leverage online environments to promote efficient digital emotion regulation as a form of transferable skill instead of didactic information delivery? How can we break down and tailor effective emotion regulation learning for an individual user?

### 3.2. Terminologies and Definitions

In this section, we describe the terms, keywords, and definitions used in our experiments and analysis.

- Tweet/Post: This is the original tweet, the source of the conversation that is being analysed. This is also the Root node in the graph that is later used to analyse the conversation and is node 1 in Figure-6 (b).

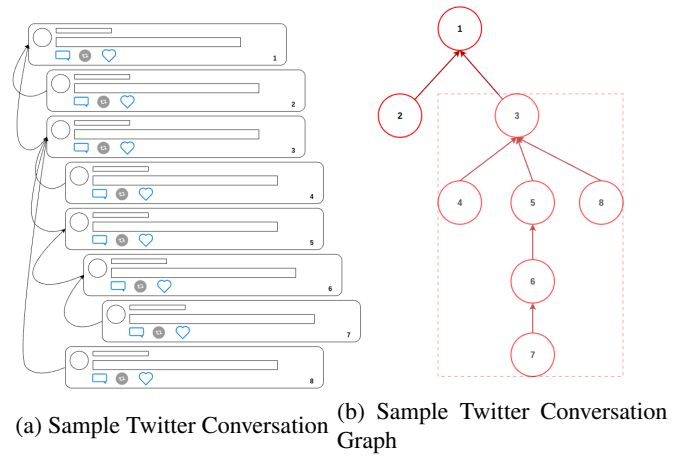

(a) Sample Twitter Conversation  (b) Sample Twitter Conversation Graph

**Figure 1:** Sample Twitter conversation (a) and a conversation graph (b) for the same. In (b) node 1 is the Root node, representing the source tweet/post, nodes 2 & 3 are comment nodes, nodes 4, 5, 6, 7, 8 are the responses and the dotted box contains the reply tree originating from node 3

- Comment(s)/Reply: A comment is a direct response to the tweet or a comment on the original post and is represented by nodes 2 & 3 in Figure-6 (b).

- Response(s): A response is a comment received on a comment, that is, it is not a direct reply to the source tweet, they are nodes 4-8 in Figure-6 (b).

- Reply tree: A reply tree is a thread generated from comments and their responses. It is represented by the dotted box in Figure-6 (b).

- Conversation: This comprises the tweet/post along with all its comments, responses and reply trees, it is represented by the graph in Figure-6 (b).

- Emotion Board: A key-value pair consisting of six elements, the keys denote the emotions and the values are a floating point number representing the cumulative proportion of each emotion exhibited by the post/tweet/root node.

- Influence: The emotion board of the root node is representative of the combination of emotional expressions of its child nodes. Therefore, every node that is not the root node, has an impact on the root node, based on its location in the graph and the emotion its text carries. The impact of a node is given by Equation-1 described in Section 3.6 of this article.

### 3.3. Methodological Framework

The eImpact framework for encouraging emotion regulation in social media conversations is depicted in Figure-2. It is composed of four key components: data retrieval, conversation emotion analysis, conversation graph analysis and identifying the context for ER. The data retrieval process starts with gathering information from social media conversations, in this case Twitter conversations. In recent years,

Twitter has been a popular destination for hashtag-based social movements such as #MeToo and #BlackLivesMatter, but the platform's free speech policy also increases the risk of hate and harassment. Therefore, we gathered a variety of Twitter conversations and saved them as CSV files. We used feature engineering to create a set of files, each containing a conversation wherein each row comprised of a text string representing a tweet or a comment, as well as their ID and metadata (parameters like the number of comments received, authors who replied etc). The second component then analyses emotion propagation using these CSV files. It begins with categorising the emotions expressed in tweets. Emotions are divided into 6 primary and 27 tertiary categories. We use the primary emotion categories in this work to classify the emotion in tweets. We then generate a graph of the conversation after classifying the tweets into 6 emotion classes. This graph is used to calculate the emotional impact of individual tweets on the entire conversation as well as the percentage distribution of various emotions in the discussion. Following that, in the final component, we use the graph to identify the nodes that have the greatest impact on the emotion of the conversation and apply this information to identify scenarios where emotion regulation needs to be undertaken and offer support for the same.

**Table 3**
Parameters used to fetch tweets

| Parameter | Description |
| --- | --- |
| author_id | The unique identifier of the user who posted the tweet/comment/response. |
| conversation_id | The unique identifier used to identify a conversation/thread on Twitter. |
| created_at | Timestamp of the tweet/comment/response in UTC. |
| id | The unique identifier of the tweet/comment/response on Twitter |
| in_reply_to_user_id | The author id of the user who received the response. |
| entities | Provides metadata and additional contextual information about Twitter posts. For instance, hashtags, user mentions, links, and so on. |
| lang | Filter used to select English tweets. |
| text | The text contained in the tweet along with the emoticons. |

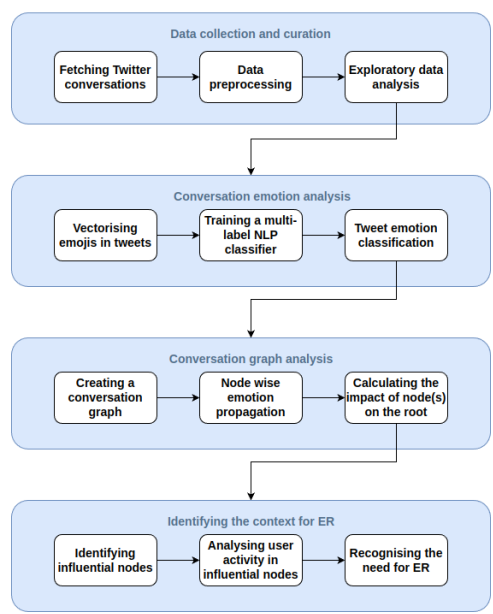

**Figure 2:** eImpact: Framework for encouraging on-spot emotion regulation in social media conversations

### 3.4. Data

Twitter is regularly used by government officials in Australia, to post updates and notify of recent events or inform citizens of upcoming activities. For the purpose of this study, we analysed the tweets by members of the Parliament, from the six Australian states, for the period between April 2020 to August 2022. The aim was to collect a variety of conversations by topic, hashtags and context. These involved tweets about the various lockdowns, COVID vaccine updates, policy updates, local developments and announcements. A total of 75 conversations were selected for this analysis, each of which involved a minimum of 1000 direct and a total of 3000 responses, leading to a dataset of 75*3000 rows. This data was downloaded using the Twitter API (Tweepy) and the tweet downloader tool provided by Twitter. Every tweet on Twitter has a conversation ID, which was used to collect and organise tweets, comments and responses. It must be noted that each of these conversations were separate tweets and not responses or quotes to another tweet. Table- 3 describes the parameters used while fetching the data.

For this experiment, tweets with only text and emoticons were taken into consideration. Tweets containing images, videos, or external links were not considered in this experiment, and the evaluation of emotions expressed in media files will be taken into account in future works. In the current data files, the tweets were mostly in English, and the occasional comments in a different language were removed. The data was downloaded in the form of CSV files, one per conversation and then used for further analysis. Users involved in a conversation were identified by their author_id and the in_reply_to_user_id was used to associate comments with its responses. The entities parameter was used to trace the sequence of responses. The number of responses to each comment and the number of unique users who responded to the comment were also added as attributes to the data.

### 3.5. Emotion Classification

Each row in the CSV file contained a text field representing either a tweet, a comment on a tweet or a response to a comment. The emotion expressed by the text in this field was determined using a text emotion classifier. A multi-label

NLP classifier was used to categorise the tweets based on the six basic emotions (love, joy, sadness, anger, fear, and surprise). The emojis in the tweets were replaced with vector representations generated by Gensim using the Emojinal library Barry, Jameel and Raza (2021), after which the tweet text was tokenized using the TweetTokenizer. The NLP classifier was trained and tested on the 'emotions' dataset, a two-column labelled dataset of Twitter messages with a text string and a label, which contains 20,000 rows of data Saravia, Liu, Huang, Wu and Chen (2018). Six emotions are described by the labels: love, joy, sadness, anger, fear, and surprise. For training, a four-layer sequential model with Bidirectional LSTM layers was used, and the data was divided 80:10:10 for training, validation, and testing. The model was trained for 20 epochs (increasing the epochs had no effect on accuracy) and achieved a testing accuracy of 87%. The trained classifier was then used to predict the emotions expressed in the Twitter conversation. Every tweet, comment, and response in the conversation thread received an emotion label and a score, with the score indicating the probability with which the classifier predicted the emotion.

### 3.6. Graph Based Emotion Propagation Analysis

In this work, we propose that a post or tweet is representative of the emotion it expresses as well as of the emotions expressed in its comments and responses. Hence, we calculate the overall emotion represented by a conversation, by summing up the emotional impact of its source tweet, comments and responses. A graph is generated to represent the analysis of the conversations. Networks from Online Social Networks (OSN) are commonly defined by a graph in which the nodes represent the users in the network and the edges represent the links between the nodes Antonakaki et al. (2021). These graphs are useful for identifying user properties such as influence, as well as network properties such as homophily. However, the goal of this study is to analyse a conversation (in which users may have participated once or multiple times) and the "impact of the users' actions" on encouraging engagement in a conversation. Rather than identifying "problematic users," the idea is to identify those "posts" that are troublesome within a conversation (or that trigger anger/hate within a conversation). The general influence or behaviour profile of users on the social network was not examined for this study. We believe that informing a user about the impact of their rude comment(s) on a particular instance, rather than ticketing them as inappropriate in general, would encourage them to self-reflect on a specific case.

It has been discovered that while public figures, media companies, and people with social influence bring a lot of initial attention to a social media post, it is the uninfluential and anonymous users who keep the engagement going and thus have a larger impact on the emotion that is propagated within a conversation Solovev and Pröllochs (2022), Mirbabaie et al. (2021), Saveski et al. (2021). Informing users of their impact on the conversation will also question their sense of anonymity on a post and serve as a motivator to respond responsibly. As a result, for this experiment, the

**Table 4**
Parameters used to find influential nodes in the conversation graph

| Parameter | Description |
| --- | --- |
| number_of_direct_responses | The in-degree of a node. |
| total_engagement_received | The number of ancestors of a node, the number of nodes in the reply tree of a node. |
| distance_from_root | The number of directed edges in the shortest path from a node to the root node. |
| page_rank | The rank of a node in the graph, based on the structure of incoming edges. |
| emotion_score | The probability with which the emotion classification model predicted the emotion for the text in the node. |

tweets, comments, and responses were treated as nodes in the graph, and the directed edges represented the nodes that received the responses. That is, if a comment has received 'n' direct responses, it will have an in-degree of 'n'. Self-loops were also present in the graph. Therefore, the conversation graph G can be defined as:

*G = (V, E, A), where V is the set of nodes, E is the set of edges representing the nodes' existing relations, and A denotes the set of attribute vectors. The value of n=|V| represents the total number of vertices, m=|E| represents the total number of edges, and A (A1, A2, A3... Ak) associates with nodes in V and describes their characteristics, where k is the number of attributes each node has.*

Finding influential nodes in a graph has been widely used in sentiment classification and the analysis of Online Social Networks (OSN). In the case of Twitter network analysis, although there is no widely accepted standard for identifying influential nodes, a combination of various connectivity/centrality-based attributes and machine learning methods has been applied Berahmand, Haghani, Rostami and Li (2020), Vilarinho and Ruiz (2018), Ban Kirigin, Bujačić Babić and Perak (2021), Bordoloi and Biswas (2020). Because global connectivity is not a significant measure of a node's impact, in this case, attributes that focus on identifying nodes with high local impact were used. This is because two separate comments on a tweet can grow into large threads regardless of their connectivity or the presence of common nodes between them. The attributes that were used to represent the nodes are described in Table-4.

The number of direct responses indicates a node's in-degree, and the total responses thread indicates a node's ancestors (number of nodes that have a path to the source). Page rank is used to rank nodes with the same in-degree; (it has been preferred over betweenness centrality in this case because it estimates the nodes' local importance Antonakaki et al. (2021)). The distance from the root is used as a factor to reinstate the distance of a trigger, a comment that initiated the reply tree is more influential than a response in the reply tree. Based on these attributes, the emotional influence ($E_m$) of child nodes and the terminating nodes for the root node was calculated using the following rule (also shown in Fig-3):

*The impact of nodes in $G = (V, E, A)$ on the root node $R$ is given by:*

$$\forall V \in G - \{R\}, E_m(R) = \sum E_m(V1, V2, V3....Vn) \quad (1)$$

where:

$$E_m(Vi) = f(Ai)$$

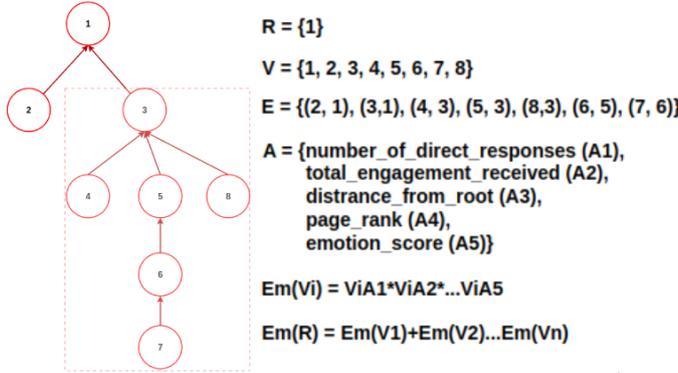

**Figure 3**: Calculating the emotion represented by the root node using emotion propagation

Node 1 is the root node, as shown in Fig-3, and it has received two direct responses (2 & 3). It also received 7 engagements (comments + responses). Nodes 2 and 3 are the closest to the root node, while node 7 is the furthest away. As a result, when calculating node 3's emotional impact on the root node, the number of direct responses would be 3, total engagement received would be 5, and distance from root would be 1 among other parameters. The rule is used to calculate the impact of nodes on a specific node, which in this case is the root. A threshold value was chosen based on the impact of nodes, which in this case was the mean value of node impacts on the root node. This was used to identify influential nodes, which were distinguished as nodes with an impact value greater than the threshold. After identifying the influential nodes $(I = \{V1, V2...Vn\})$ for the root node $(R)$, the same rule can be used to find the nodes with the greatest impact on these influential nodes $(I_{V1} = \{V1_1, V1_2...V1_n\})$ by considering the subgraph where the influential nodes $(V1, V2...Vn)$ are the root nodes.

### 3.7. Proposed Framework

To estimate the nodes that have the greatest impact on the root node, we combine the influential nodes identified by eImpact with those with a high toxicity score indicated by the Perspective API Google (2021). The Perspective API predicts the perceived impact of a comment on a conversation by analysing it across a variety of emotional concepts. It returns a probability score (a value in the range of 0-1). The score represents the percentage of readers who consider the comment to be toxic. To ensure consistency across our experiments, we calculated a toxicity threshold value (0.9) based on the mean value reported in the analysis. The Perspective API Google (2021) also recommends this threshold value for research; a high value reduces the possibility of bias. A higher score indicates that a reader is more likely to perceive the comment as containing the given attribute. The Perspective API currently does not take the context of the comment into account. The intersection of the most influential nodes identified by the eImpact framework and the highly toxic nodes indicated by the toxicity scores is used.

### 3.8. Supporting On-Spot Emotion Regulation

Conversations in the form of comments or replies to posts are a pertinent aspect of social media platforms. However, they also increase the likelihood of online hate and harassment. According to research, 41% of US adults have been victims of online hate and harassment Thomas et al. (2022). Hence, this work aims to break down the occurrence of hate speech in conversations by quantifying their impact on the overall conversation, so that the users can be informed about their micro-impact on a conversation and be encouraged to act responsibly.

After identifying the influential nodes in the conversation graph, the information can be used to inform users involved in the comment node's reply tree about the implications of their actions. This can be accompanied by highlighting the node based on the colour of the emotion it represents, as well as the percentage distributions of emotions, or by restricting further activity on the comment that is the reply tree's originating node. As shown in Figure-4, the post has colour-coded comments and associated reply trees in conversations according to the emotions they represent, the distribution of those emotions, and the influential nodes that have a significant impact on that distribution. Nodes 3, 5 and 6 are particularly key in the conversation's anger, with node 6 having the biggest contribution. As a result, it is frozen.

The decision to freeze a comment here is an attempt to identify the actions of a user or a group of users in order to avoid the post becoming hostile as a whole, as well as to decompose the toxic activity involved in a post by bringing it to light. It should be noted that suppression of anger or hate speech in this context is not the same as the inherent bias of promoting positive content on social media, but rather an attempt to avoid the induction of excessive hate on a post because it's simple to respond to a digitally induced emotional challenge with a digital action, such as expressing anger on a post by leaving an angry comment Wang and Diakopoulos (2021), Smith et al. (2022). The first step is self-reflection, which is similar to cognitive and dialectic behaviour therapy, which emphasises emotion regulation. Encouraging self-reflection by outlining the effects of previous user actions, will lead to a rise in awareness among the users and nudge them towards acting responsibly.

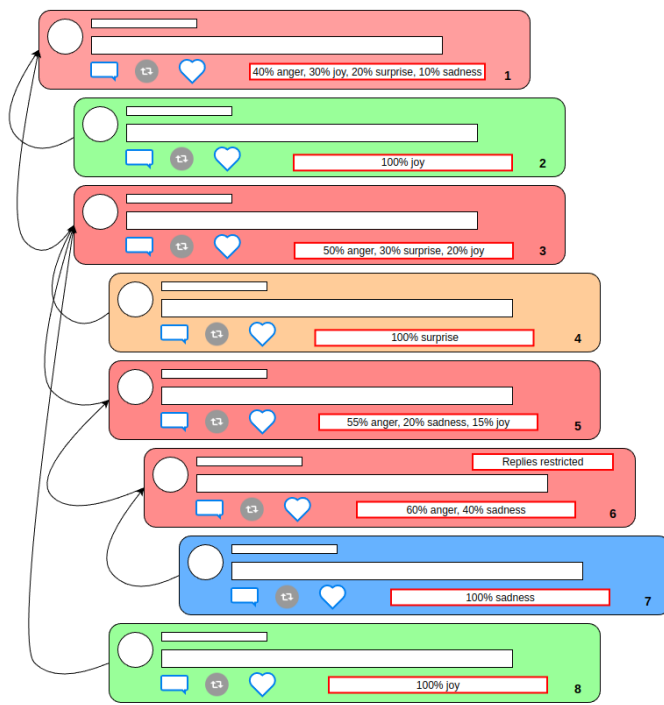

**Figure 4:** Identifying influential comments/responses/threads in social media conversations. The comments are colour-coded according to the emotion they represent.

Implicit emotion control strategies involving cognitive change, such as reappraisal or non-judgemental acceptance, can be recommended to assist the users affected by the restricted activity on their comments and prevent the creation of similar reply trees. This can be fuelled by informing users about the overall emotional distribution of the conversation in an effort to help them empathise with other participants, which will subtly prompt the user to feel responsible. This approach avoids the users from falling into the trap of anxiety based on their general perception while their anonymity is in question since it focuses on the actions of the users rather than the users themselves. Although it is debatable if preventing additional activity on a comment creates friction for participation, it protects users' right to free speech while posing a minor distraction, and the users' freedom of choice still remains with them Kiskola et al. (2021). Additionally, it has been found that one of the best ways to reduce toxicity online is through the moderation of online content Thomas et al. (2022), Jhaver et al. (2021). Questioning the users' anonymity by displaying their impact on a conversation and its emotion distribution, or freezing a particular comment in a post (depending on the length of the post, the number of persons involved, or the intensity of the emotions) serves as a modest warning of rising toxicity in this situation.

## 4. Experimental Evaluation and Analysis of Emotion Propagation

Toxicity in social media platforms has been observed in the form of hate speech, harassment, trolling, and cyberbullying, to name a few. Although these harmful activities have different intentions, they all follow the same path: uncivilised behaviour, or, in the case of text-based social media platforms, uncivil language. Recent research has looked into how Twitter conversations can be used to detect harassment and cyberbullying Guberman et al. (2016), Georgakopoulos, Tasoulis, Vrahatis and Plagianakos (2019), Pavlopoulos, Sorensen, Dixon, Thain and Androutsopoulos (2020). Not only do these require contextual information and user profile data, but they are usually applied to a conversation after it has occurred. Social media platforms have implemented moderation strategies such as deplatforming and removing inappropriate comments, but they are based on static rules and are typically applied after the incident has occurred Jhaver et al. (2021). We introduce a framework for studying the transmission of emotions (via language) in real-time online conversations. The potential of this approach lies not only in determining whether or not a conversation is becoming toxic but also in determining why and how this is happening. We believe this framework will help reduce the possibility of a conversation becoming toxic by investigating the actions that lead to toxicity.

On social media platforms where uncivil language is commonly used by a large number of users, hate speech-based toxicity has been discovered to be an important component of virality Maarouf et al. (2022), Goel et al. (2016). A viral spread corresponds to a rapid dissemination of a piece of information via connective action of users and is often fuelled by the emotions it entails. According to studies, both hate speech and toxicity spiral in action, resulting in a faster and more widespread dispersal. As a result, by monitoring the nature of information diffusion as well as the compounding effect of emotion propagation within a conversation graph, it is possible to identify when and how a conversation may become toxic and thus apply moderation strategies before it becomes toxic. The same method can be used to assess the context when emotion regulation is needed. The goal here is to inform users involved in a conversation about the emotional impact of their actions and to gently nudge them towards a responsible mindset through repeated application of the same.

A toxic tweet on Twitter is defined as an unreasonably rude or disrespectful comment that causes a person to leave the conversation. Hence, for this work, we define toxicity in a conversation as the disproportionate or excessive influx of toxic tweets/responses into an otherwise neutral post Aroyo, Dixon, Thain, Redfield and Rosen (2019), Xenos, Pavlopoulos, Androutsopoulos, Dixon, Sorensen and Laugier (2021). As previously discussed (Section 3.5) in this paper, we categorise tweets into six emotions (namely, anger, love, joy, fear, sadness, and surprise). The influx of anger has been regarded as rage-inducing, rude, or disrespectful, and is thus regarded as an indicator of toxicity.

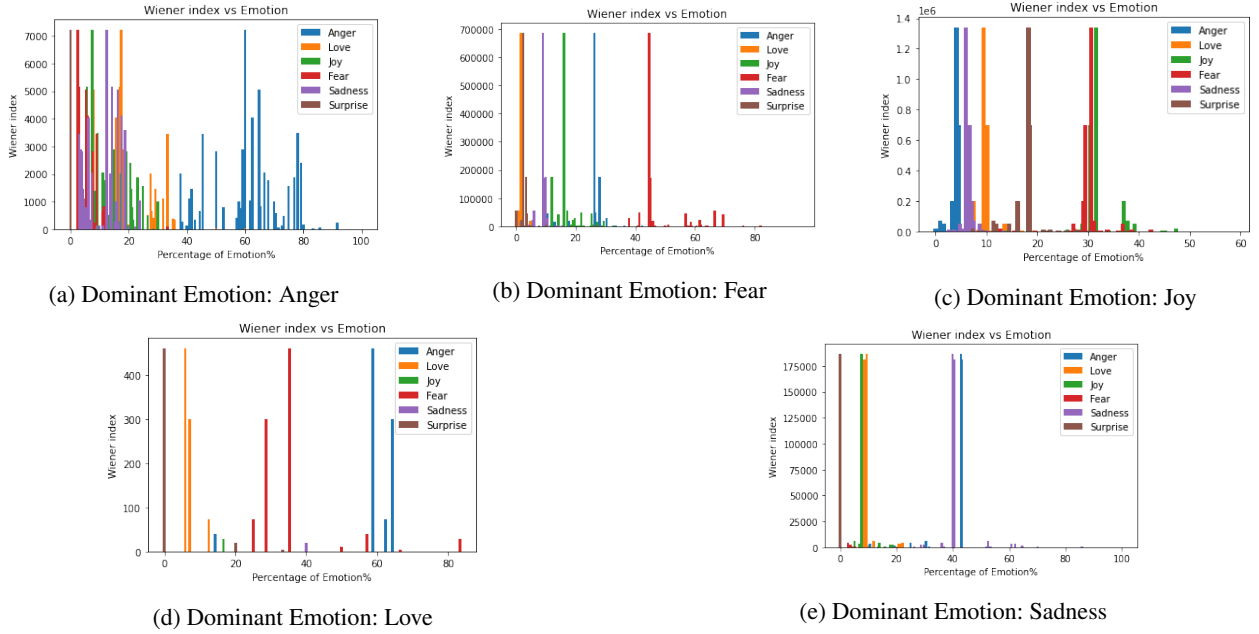

Figure 5: Wiener Index of the influential nodes with respect to the percentage of emotions in the reply tree where the dominant emotion is (a) Anger (b) Fear (c) Joy (d) Love and (e) Sadness

The structure and toxicity of Twitter conversations have been extensively researched. We use the findings from literature to validate the applicability of our framework. We do so based on three attributes: structural characteristics, emotions and virality, and the development of connective action.

### 4.1. Structural Characteristics

Users can respond to a tweet/post by liking, sharing, or commenting on it. These actions may elicit additional responses from other users. This results in a conversation graph with reply trees that are rooted in the original tweet. Toxicity has been shown to increase as the size, density, and width of a conversation graph increase Kanavos, Perikos, Vikatos, Hatzilygeroudis, Makris and Tsakalidis (2014), Saveski et al. (2021). Table-5 shows the average distribution of emotions and the number of unique users involved in influential node reply trees for each emotion represented by the influential node whereas Table-6 displays the average difference (increase/decrease) in the distribution of emotions in the graph after calculating emotional propagation using the eImpact framework for each emotion represented by the influential node, as well as the initial percentage of the represented emotion. As shown in Table 6, as the size of the conversation grows, it becomes dominated by a single emotion that has been expressed the most, as opposed to the initial distribution of emotions. If we consider a reply tree's toxicity to be the fraction of outrageous tweets it receives, we can see that it increases with the size, width, and density of the reply tree. The percentage of angry tweet responses skyrockets at 40% and continues to rise until it reaches 80% (Figure-5 (a)). Because anger spreads faster than other emotions, the same pattern can be seen when comparing the percentage of emotions contained in the nodes to the total number of nodes in the graph (Table-6).

Recently, the Wiener index has been used to analyse the structure of information diffusion, specifically to determine whether information spreads in a broadcast or viral manner Saveski et al. (2021), Goel et al. (2016). The Wiener index w(T) of a response tree T is defined as the average distance between all pairs of nodes in the tree, as given by Equation 2. A low Wiener index is associated with the broadcast structure of diffusion, in which users mostly respond to the original tweet and there is little engagement among users, whereas a high Wiener index value is associated with viral spread, in which back-and-forth engagement among users is greater than the responses to the original tweet.

$$w(T) = 1/n(n-1) \sum_{i=1}^{n} \sum_{j=1}^{n} d_{ij}, \qquad (2)$$

where:

$d_{ij}$ = the length of the shortest path between nodes i and j

After identifying the conversation graph's influential nodes, we calculated their Wiener index with respect to the percentage of emotions contained in their reply trees. Figure-5 shows the influential node's Wiener index (y-axis) versus the percentage (x-axis) of various emotions in the influential node's reply tree for each emotion. It was discovered that the Wiener index for anger tends to increase the most. It begins low and starts to rise as the percentage of anger in the reply tree increases. It peaks at 60% anger in the reply tree nodes and then drops by a small value before plateauing

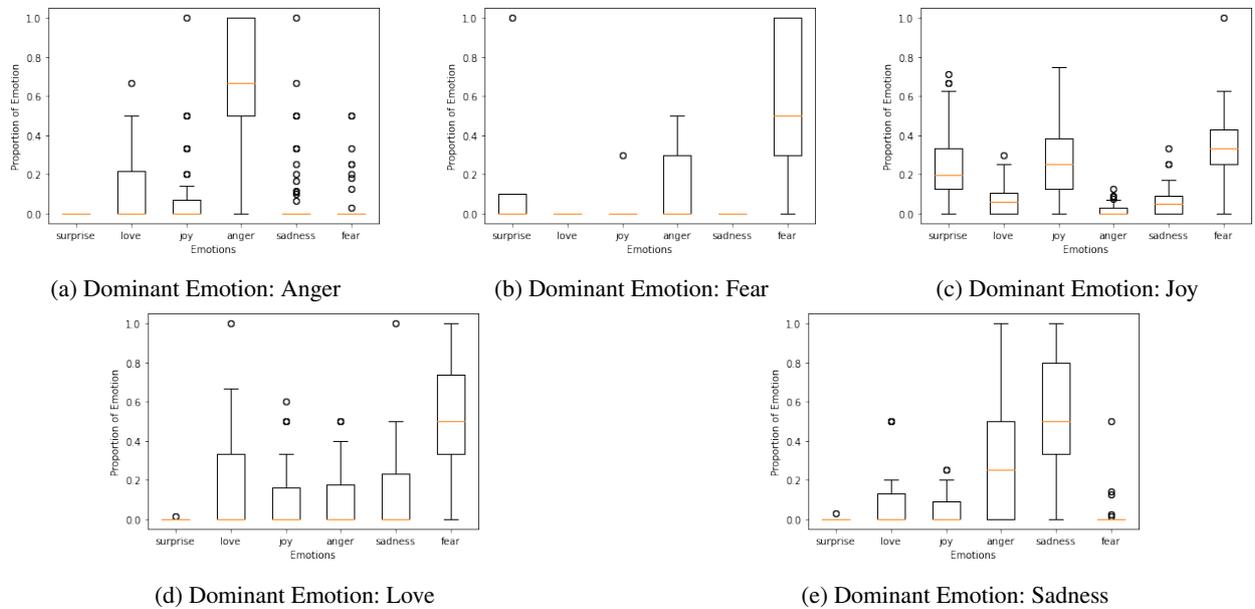

Figure 6: Percentage distribution of emotions in the reply tree of influential nodes of conversations where the dominant emotion is (a) Anger (b) Fear (c) Joy (d) Love and (e) Sadness

at the end, as shown in Figure-5 (a). This leads us to the conclusion that the influential node initially broadcasts a variety of emotions, and as the size of its reply tree and the percentage of anger grows, it transforms into a viral spread in which the nodes of this reply tree receive more back-and-forth engagement. Thereafter, once the percentage of anger becomes 60%, due to the back-and-forth engagement in multiple nodes of this reply tree, the nodes start broadcasting the same emotion and hence pivot into a toxic conversation as the conversation is now concentrated in these few nodes leading to the propagation of anger. Although the pattern of the conversation pivoting from broadcast to viral and back to broadcasting can be seen across other emotions, as shown in Figure-5, the peaks converge much sooner (35-40%) and at a higher value (10k) than anger (b, c, d and e). It emphasises the fact that anger spreads faster by convergent virality and earlier broadcasting. It is also clear that negative sentiments converge faster than positive ones. This could be because positive emotional expressions are bombarded with negativity, causing a shift in the distribution of emotions in the reply tree and a gradual shift in the form of transmission. As can be seen from Table-6, the average responses in a reply tree for different emotions. These findings are consistent with those found in the literature, thereby validating the applicability of our hypothesis.

### 4.2. Emotions and Virality

The success of online content nowadays relies upon how well it can influence others. This is also the motivation for online posts, as well as a factor in expanding one's reach. When the user activity in the influential nodes was examined, it was discovered that the number of responses received by a tweet is directly proportional to the amount of emotion contained in the tweet. Although tweets expressing anger and love received the most responses, the majority of these responses were angry. Furthermore, tweets expressing love received the angriest or rage-inducing responses. As shown in Fig-5 (a), when the influential node has anger as the highest percentage of expression, the initial distribution of emotions represented by the nodes changes as the reply tree multiplies, and anger begins to dominate, reaching 70-85% of the overall expression in most cases. When a node strongly expresses love, the response tree branches into anger as well. Love receives a large number of polarising and rage-inducing responses (Table-5), which causes anger to dominate the emotional expression in the reply tree, as shown in Fig-5 (d).

**Table 5**
Average distribution of emotions and the number of unique users in the reply trees of influential nodes

| Emotion represented by influential node | Average responses to an influential node based on its emotion (%) | | | | | No. of unique users |
|---|---|---|---|---|---|---|
| | Anger | Fear | Love | Joy | Sadness | |
| Anger | 78.6 | 4.2 | 5.3 | 4.7 | 5.4 | 3 |
| Fear | 31.3 | 56.8 | 4.2 | 1.3 | 6.4 | 5 |
| Joy | 37.4 | 1.4 | 15.7 | 43.1 | 2.3 | 7 |
| Love | 81.3 | 1.2 | 8.9 | 6.7 | 1.6 | 7 |
| Sadness | 32.8 | 13.2 | 8.7 | 2.4 | 42.6 | 4 |

**Table 6**
Average change in the distribution of emotions in the graph after calculating the propagation of emotions using the eImpact framework

| Emotion represented by the root node | Average change in the distribution of emotions (%) | | | | | Initial % of nodes for the emotion | No. of nodes in the graph |
|---|---|---|---|---|---|---|---|
| | Anger | Fear | Joy | Love | Sadness | | |
| Anger | 27.6 | -7.4 | -12.8 | -13.4 | -4.1 | 38.3 | 3343 |
| Fear | -5.3 | 20.1 | -10.6 | -4.2 | 0.4 | 21.9 | 5493 |
| Joy | -3.2 | 4.2 | 15.4 | -2.7 | 0.6 | 23.4 | 5073 |
| Love | 1.7 | 2.8 | 0.3 | 12.5 | -3.4 | 20.8 | 4457 |
| Sadness | -2.1 | 2.3 | -2.4 | -10.1 | 21.2 | 26.5 | 3623 |

Although the percentage of anger expressed in cases of other emotions is not the same, the virality of anger matches the highest magnitude in each case, albeit at a relatively low percentage. Figure-6 represents the percentage distribution (y-axis, ranging from 0-1) of various emotions in the reply tree of the influential node for each emotion. The percentage distribution of emotions in the reply trees of influential nodes, as shown in Fig-6, demonstrates the same point about anger expressed in responses. Only in the case of joy does anger stay minimal. When anger is the dominant emotion, joy and fear are present in similar proportions, but they have an inverse relationship otherwise. Fig-5 (e) shows that, while sadness retains its ability to broadcast, it also receives a significant amount of anger, as shown in Fig-6 (e).

Further analysis showed that these responses were the result of repetitive responses from the same group of users. It was discovered that these responses had a polarising effect, which seemed to be exacerbated by the repeated participation of users who identified with the ideology, eventually creating hostility toward the expression of any other emotion. A similar polarisation has been observed in the literature, where other-condemning emotions embedded in tweets initiate the rumour cascade and may act as accelerators for its spread. This implies that when radical ideas and beliefs expressed in a conversation become entrenched, are more likely to manifest as toxicity. When ideological polarisation occurs, an aggressive mix of fury emotions accelerates their spread on social media conversations Solovev and Pröllochs (2022).

Table-6 also shows that, when comparing anger to other emotions, the average increase in the percentage of nodes exhibiting anger after the calculation of emotion propagation is the maximum. The ratio of angry responses received by an angry comment to comments expressing any emotion other than love also reveals the same. This was true for all cases in our study as well as for all emotions. Table-6 depicts the percentage increase in emotion represented by the nodes in the conversation graph following propagation. Because anger is known to spread more quickly, it not only occurs in the greatest proportion when it is the dominant emotion, but it also multiplies the most even when only a few users are involved. It is also in cases of intense anger/rage expression that emotions such as love and joy are most nullified (Table-6). When comparing the cases of anger and other emotions, the same pattern can be seen in the number of responses from a single user as shown in Table-5.

### 4.3. Development of Connective Action

Social media has evolved into a platform for social movements that fall under the umbrella of connective action. User interactions with posts are what initiate and sustain post activity. The activity is maintained by anonymous users Mirbabaie et al. (2021), Saveski et al. (2021). The development of connective action can be observed from Tables-5&6, where with the involvement of as few as 3 users in various reply trees (number of nodes in the reply trees range from 112-270) changes the emotion distribution of a graph containing more than 3000 nodes by 4-27%. It has been discovered that toxicity on a post is caused by the unimportant and unrelated contents or comments. Journalists, celebrities, and other public figures may help a tweet gain traction, but they are more likely to post promotional content. Private individuals or general users not only sustain a post's engagement but also incite hate speech or toxicity. This could be because the primary goal of general users is to use the platform to express and connect with others. Given that they are not the most retweeted or popular, emotional expression becomes their primary means of communicating their ideas to the public. As a result, their anonymous actions become the impetus for a post which is reflected in Table-5&6 by the number of unique users involved in the responses to a node. An average of 3 unique users can flip the dispersal of a node from broadcast to viral and accumulate up to 27% increase in anger but even 7 unique users on average are unable to shift the emotion distribution of the node by half of this value when the emotion being expressed is love or joy. This also fits with the concept of emotional mobilisation, which is said to be necessary for launching a social movement among the psychological states of a group of people Castells (2015), Saveski et al. (2021). The tweet that started the #MeToo movement on Twitter included both necessary outrage and hope for possible change.

**Table 7**
Comparison of proposed framework against Perspective API Google (2021)

| Utilised Model | Identification of influential nodes | Percentage of nodes identified as toxic | Possible reduction in toxicity if acted on these nodes |
|---|---|---|---|
| eImpact Framework | Based on impact score, taking into account the tweet text as well as its connectivity | 1-4% | 10% |
| Perspective API Google (2021) | Based on toxicity, taking into account only the tweet text | 1-2% | 7% |
| Proposed Framework | Takes into account the emotion in the tweet text, the connectivity as well as its toxicity | 1-4% | 12% |

This necessitates dealing with the imbalanced induction of anger on social media posts in a sensitive and non-suppressive manner. Deplatforming or barring a user from making further comments may temporarily reduce the amount of anger on a post, but it does little in the long run because the root cause of irresponsible behaviour is not addressed. Since widespread toxicity is caused by collective action, analysing the propagation of emotions based solely on user actions rather than users, in general, will help reduce toxicity-causing actions. As a result, diluting users' anonymity by exposing the consequences of their actions will encourage implicit emotion regulation, provoking self-reflection and responsible behaviour.

### 4.4. Performance Evaluation

Aside from preventing the compounding of anger or hate speech in general that later becomes toxic, the proposed framework considers the subjectivity of toxicity by not relying solely on the content of a post/reply/comment and considering its impact on the ongoing conversation. We evaluated this framework by comparing it to the toxicity scores generated by the Google Perspective API Google (2021). Following that, we classified a tweet/comment as toxic based on this threshold value. When the toxicity of the conversation was examined, it was discovered that the most toxic tweets/comments were among the most influential nodes. It was also discovered that the reply tree of influential nodes contained more than 87% of the overall toxicity of the conversation. We used the toxicity score from the Perspective API Google (2021) to find the influential nodes because the goal is to reduce the induction of hate speech in a conversation by analysing its influential nodes. Thereafter, we used a combination of the eImpact results and toxicity scores as the proposed framework to identify influential nodes in a conversation. We discovered that while the influential nodes generated by the eImpact framework and the Perspective API have a 75% match, the influential nodes generated by the proposed framework and the toxicity scores have a 94.3% match. This confirms that the proposed framework can be used to identify comments that may be contributing to the toxicity of a post while taking into account their subjectivity and whether or not they are individually toxic. This framework can be used in conjunction with online content moderation to understand the source(s) of toxicity in conversations by taking both content and context into account. Table-7 compares the proposed framework to the Perspective API in terms of identifying influential nodes and the possibility of toxicity reduction. As shown in Table-7, the proposed framework improves the identification of toxic tweets by 3% and opens up the possibility of reducing toxicity by 5%. This increase is due to the use of both context and content-based node evaluation in the conversation graph.

We also looked at how much hate speech can be reduced by deactivating influential nodes or restricting further activity on these nodes once they reach the toxicity threshold. We discovered that, while the eImpact framework could reduce hate speech by 10% and the Perspective API by 7%, the proposed framework which is a combination of the eImpact framework and the toxicity scores from the Perspective API Google (2021) could reduce toxicity by 12%. This is consistent with the understanding that restricting activity on a comment may not prevent users from commenting elsewhere or starting a new thread on the same post; however, because the restriction is accompanied by factual information about one's impact on the post, we believe it will encourage self-reflection and implicit emotion regulation.

## 5. Conclusion

Researchers have addressed concerns about the negative effects of online toxicity on mental health. Although knowledge of digital ER may not eliminate these concerns, it does imply that some digital media may be used in conjunction with constructive psychological techniques for significant instrumental purposes, such as improving work performance or fostering social harmony. Because text-based discussions and comment threads are such an important part of online communication, they frequently serve as breeding grounds for negativity. Understanding how to identify the context when emotional regulation is needed and supporting users by providing opportunities for self-reflection and emotional regulation may substantially reduce the generation of rage-inducing posts and improve online well-being. In this paper, we propose a graph-based framework to identify such posts in the form of influential nodes in a conversation graph. We focused on the structural and contextual elements of a conversation by performing a micro-level analysis. We tested the utility of the proposed framework by comparing it with the Perspective API and described how the framework can be used to recede toxicity in conversations. These findings contribute to our understanding of how emotion dysregulation can lead to toxic behaviour online and how support for effective emotion regulation can mitigate this. This framework can be used to inform the design of healthier social media conversation/discussion platforms, specifically to maintain the quality of discussions, which has been acknowledged as an ongoing challenge in online digital technology.

## References

Antonakaki, D., Fragopoulou, P., Ioannidis, S., 2021. A survey of twitter research: Data model, graph structure, sentiment analysis and attacks. Expert Systems with Applications 164, 114006.

Aroyo, L., Dixon, L., Thain, N., Redfield, O., Rosen, R., 2019. Crowdsourcing subjective tasks: the case study of understanding toxicity in online discussions, in: Companion proceedings of the 2019 world wide web conference, pp. 1100–1105.

Ban Kirigin, T., Bujačić Babić, S., Perak, B., 2021. Lexical sense labeling and sentiment potential analysis using corpus-based dependency graph. Mathematics 9, 1449.

Barry, E., Jameel, S., Raza, H., 2021. Emojional: Emoji embeddings, in: UK Workshop on Computational Intelligence, Springer. pp. 312–324.

Berahmand, K., Haghani, S., Rostami, M., Li, Y., 2020. A new attributed graph clustering by using label propagation in complex networks. Journal of King Saud University-Computer and Information Sciences .

Bordoloi, M., Biswas, S.K., 2020. Graph based sentiment analysis using keyword rank based polarity assignment. Multimedia Tools and Applications 79, 36033–36062.

Brambilla, M., Javadian, A., Sulistiawati, A.E., 2021. Conversation graphs in online social media, in: International Conference on Web Engineering, Springer. pp. 97–112.

Castells, M., 2015. Networks of outrage and hope: Social movements in the Internet age. John Wiley & Sons.

Chen, Y., Mark, G., Ali, S., 2016. Promoting positive affect through smartphone photography. Psychology of well-being 6, 1–16.

Chowanda, A., Sutoyo, R., Tanachutiwat, S., et al., 2021. Exploring text-based emotions recognition machine learning techniques on social media conversation. Procedia Computer Science 179, 821–828.

Emotion Regulation in Social Media Conversations

Costa, J., Guimbretière, F., Jung, M.F., Choudhury, T., 2019. Boostmeup: Improving cognitive performance in the moment by unobtrusively regulating emotions with a smartwatch. Proceedings of the ACM on Interactive, Mobile, Wearable and Ubiquitous Technologies 3, 1–23.

Dmitrenko, D., Maggioni, E., Brianza, G., Holthausen, B.E., Walker, B.N., Obrist, M., 2020. Caroma therapy: pleasant scents promote safer driving, better mood, and improved well-being in angry drivers, in: Proceedings of the 2020 chi conference on human factors in computing systems, pp. 1–13.

Georgakopoulos, S.V., Tasoulis, S.K., Vrahatis, A.G., Plagianakos, V.P., 2019. Convolutional neural networks for twitter text toxicity analysis, in: INNS Big Data and Deep Learning conference, Springer. pp. 370–379.

Gillespie, T., 2020. Content moderation, ai, and the question of scale. Big Data & Society 7, 2053951720943234.

Goel, S., Anderson, A., Hofman, J., Watts, D.J., 2016. The structural virality of online diffusion. Management Science 62, 180–196.

Goldenberg, A., Gross, J.J., 2020. Digital emotion contagion. Trends in Cognitive Sciences 24, 316–328.

Google, 2021. Perspective API. https://www.perspectiveapi.com/. [Online; accessed 19-Dec-2022].

Gorwa, R., Binns, R., Katzenbach, C., 2020. Algorithmic content moderation: Technical and political challenges in the automation of platform governance. Big Data & Society 7, 2053951719897945.

Gross, J.J., 2014. Emotion regulation: conceptual and empirical foundations. .

Gross, J.J., 2015. Emotion regulation: Current status and future prospects. Psychological inquiry 26, 1–26.

Guberman, J., Schmitz, C., Hemphill, L., 2016. Quantifying toxicity and verbal violence on twitter, in: Proceedings of the 19th ACM Conference on Computer Supported Cooperative Work and Social Computing Companion, pp. 277–280.

Hossain, E., Wadley, G., Berthouze, N., Cox, A., 2022. Motivational and situational aspects of active and passive social media breaks may explain the difference between recovery and procrastination, in: CHI Conference on Human Factors in Computing Systems Extended Abstracts, pp. 1–8.

Jhaver, S., Boylston, C., Yang, D., Bruckman, A., 2021. Evaluating the effectiveness of deplatforming as a moderation strategy on twitter. Proceedings of the ACM on Human-Computer Interaction 5, 1–30.

Kanavos, A., Perikos, I., Vikatos, P., Hatzilygeroudis, I., Makris, C., Tsakalidis, A., 2014. Conversation emotional modeling in social networks, in: 2014 IEEE 26th International Conference on Tools with Artificial Intelligence, IEEE. pp. 478–484.

Kiskola, J., Olsson, T., Väätäjä, H., H. Syrjämäki, A., Rantasila, A., Isokoski, P., Ilves, M., Surakka, V., 2021. Applying critical voice in design of user interfaces for supporting self-reflection and emotion regulation in online news commenting, in: Proceedings of the 2021 CHI conference on human factors in computing systems, pp. 1–13.

Konikoff, D., 2021. Gatekeepers of toxicity: Reconceptualizing twitter's abuse and hate speech policies. Policy & Internet .

Kou, Y., Gui, X., 2020. Emotion regulation in esports gaming: a qualitative study of league of legends. Proceedings of the ACM on Human-Computer Interaction 4, 1–25.

Li, S., Hao, Y., Yoon, J., 2020. Purpal: An interactive box that up-regulates positive emotions in consumption behaviors, in: Extended Abstracts of the 2020 CHI Conference on Human Factors in Computing Systems, pp. 1–6.

Lukoff, K., Yu, C., Kientz, J., Hiniker, A., 2018. What makes smartphone use meaningful or meaningless? Proceedings of the ACM on Interactive, Mobile, Wearable and Ubiquitous Technologies 2, 1–26.

Luo, Ruikun, N.D., Yang, X.J., 2021. Behavioral and physiological signals-based deep multimodal approach for mobile emotion recognition. IEEE Transactions on Affective Computing .

Maarouf, A., Pröllochs, N., Feuerriegel, S., 2022. The virality of hate speech on social media. arXiv preprint arXiv:2210.13770 .

Majó-Vázquez, S., Nielsen, R., Verdú, J., Rao, N., de Domenico, N., Papaspiliopoulos, O., 2020. Volume and patterns of toxicity in social media conversations during the covid-19 pandemic .

Majumder, N., Poria, S., Hazarika, D., Mihalcea, R., Gelbukh, A., Cambria, E., 2019. Dialoguernn: An attentive rnn for emotion detection in conversations, in: Proceedings of the AAAI conference on artificial intelligence, pp. 6818–6825.

Mancini, M., Niewiadomski, R., Huisman, G., Bruijnes, M., Gallagher, C.P., 2020. Room for one more?-introducing artificial commensal companions, in: Extended Abstracts of the 2020 CHI Conference on Human Factors in Computing Systems, pp. 1–8.

McRae, K., Gross, J.J., 2020. Emotion regulation. Emotion 20, 1.

Ments, L.v., Treur, J., Klein, J., Roelofsma, P., 2021. A second-order adaptive network model for shared mental models in hospital teamwork, in: International Conference on Computational Collective Intelligence, Springer. pp. 126–140.

Mirbabaie, M., Brünker, F., Wischnewski, M., Meinert, J., 2021. The development of connective action during social movements on social media. ACM Transactions on Social Computing 4, 1–21.

Miri, P., Jusuf, E., Uusberg, A., Margarit, H., Flory, R., Isbister, K., Marzullo, K., Gross, J.J., 2020. Evaluating a personalizable, inconspicuous vibrotactile (piv) breathing pacer for in-the-moment affect regulation, in: Proceedings of the 2020 CHI Conference on Human Factors in Computing Systems, pp. 1–12.

Mohammad, S.M., Turney, P.D., 2013. Crowdsourcing a word–emotion association lexicon. Computational intelligence 29, 436–465.

Newbold, J.W., Luton, J., Cox, A.L., Gould, S.J., 2017. Using nature-based soundscapes to support task performance and mood, in: Proceedings of the 2017 CHI Conference Extended Abstracts on Human Factors in Computing Systems, pp. 2802–2809.

Paredes, P.E., Zhou, Y., Hamdan, N.A.H., Balters, S., Murnane, E., Ju, W., Landay, J.A., 2018. Just breathe: In-car interventions for guided slow breathing. Proceedings of the ACM on Interactive, Mobile, Wearable and Ubiquitous Technologies 2, 1–23.

Pavlopoulos, J., Sorensen, J., Dixon, L., Thain, N., Androutsopoulos, I., 2020. Toxicity detection: Does context really matter? arXiv preprint arXiv:2006.00998 .

Perikos, I., Hatzilygeroudis, I., 2018. A framework for analyzing big social data and modelling emotions in social media, in: 2018 IEEE Fourth International Conference on Big Data Computing Service and Applications (BigDataService), IEEE. pp. 80–84.

Poria, S., Majumder, N., Mihalcea, R., Hovy, E., 2019. Emotion recognition in conversation: Research challenges, datasets, and recent advances. IEEE Access 7, 100943–100953.

Ruensuk, M., Cheon, E., Hong, H., Oakley, I., 2020. How do you feel online: Exploiting smartphone sensors to detect transitory emotions during social media use. Proceedings of the ACM on Interactive, Mobile, Wearable and Ubiquitous Technologies 4, 1–32.

Ruvimova, A., Kim, J., Fritz, T., Hancock, M., Shepherd, D.C., 2020. " transport me away": Fostering flow in open offices through virtual reality, in: Proceedings of the 2020 CHI Conference on Human Factors in Computing Systems, pp. 1–14.

Saravia, E., Liu, H.C.T., Huang, Y.H., Wu, J., Chen, Y.S., 2018. CARER: Contextualized affect representations for emotion recognition, in: Proceedings of the 2018 Conference on Empirical Methods in Natural Language Processing, Association for Computational Linguistics, Brussels, Belgium. pp. 3687–3697. URL: https://www.aclweb.org/anthology/D18-1404, doi:10.18653/v1/D18-1404.

Saveski, M., Roy, B., Roy, D., 2021. The structure of toxic conversations on twitter, in: Proceedings of the Web Conference 2021, pp. 1086–1097.

Semertzidis, N., Scary, M., Andres, J., Dwivedi, B., Kulwe, Y.C., Zambetta, F., Mueller, F.F., 2020. Neo-noumena: Augmenting emotion communication, in: Proceedings of the 2020 CHI conference on human factors in computing systems, pp. 1–13.

Shaver, P., Schwartz, J., Kirson, D., O'connor, C., 1987. Emotion knowledge: further exploration of a prototype approach. Journal of personality and social psychology 52, 1061.

Shen, K., Cox, A., 2020. Video games as a tool for digital emotion regulation. HCI-E MSc Final Project Report 2020 .

Shi, Y., Koval, P., Kostakos, V., Goncalves, J., Wadley, G., 2023. "instant happiness": Smartphones as tools for everyday emotion regulation.

International Journal of Human-Computer Studies 170, 102958.

Slovak, P., Antle, A.N., Theofanopoulou, N., Roquet, C.D., Gross, J.J., Isbister, K., 2022. Designing for emotion regulation interventions: an agenda for hci theory and research. arXiv preprint arXiv:2204.00118 .

Smith, W., Wadley, G., Webber, S., Tag, B., Kostakos, V., Koval, P., Gross, J.J., 2022. Digital emotion regulation in everyday life, in: CHI Conference on Human Factors in Computing Systems, pp. 1–15.

Solovev, K., Pröllochs, N., 2022. Moral emotions shape the virality of covid-19 misinformation on social media, in: Proceedings of the ACM Web Conference 2022, pp. 3706–3717.

Steinert, S., 2021. Corona and value change. the role of social media and emotional contagion. Ethics and Information Technology 23, 59–68.

Steinert, S., Dennis, M.J., 2022. Emotions and digital well-being: on social media's emotional affordances. Philosophy & Technology 35, 1–21.

Tag, B., van Berkel, N., Vargo, A.W., Sarsenbayeva, Z., Colasante, T., Wadley, G., Webber, S., Smith, W., Koval, P., Hollenstein, T., et al., 2022a. Impact of the global pandemic upon young people's use of technology for emotion regulation. Computers in Human Behavior Reports 6, 100192.

Tag, B., Sarsenbayeva, Z., Cox, A.L., Wadley, G., Goncalves, J., Kostakos, V., 2022b. Emotion trajectories in smartphone use: Towards recognizing emotion regulation in-the-wild. International Journal of Human-Computer Studies 166, 102872.

Theofanopoulou, N., Isbister, K., Edbrooke-Childs, J., Slovák, P., et al., 2019. A smart toy intervention to promote emotion regulation in middle childhood: Feasibility study. JMIR mental health 6, e14029.

Thomas, K., Kelley, P.G., Consolvo, S., Samermit, P., Bursztein, E., 2022. "it's common and a part of being a content creator": Understanding how creators experience and cope with hate and harassment online, in: CHI Conference on Human Factors in Computing Systems, pp. 1–15.

Torre, J.B., Lieberman, M.D., 2018. Putting feelings into words: Affect labeling as implicit emotion regulation. Emotion Review 10, 116–124.

Vilarinho, G.N., Ruiz, E.E.S., 2018. Global centrality measures in word graphs for twitter sentiment analysis, in: 2018 7th Brazilian Conference on Intelligent Systems (BRACIS), IEEE. pp. 55–60.

Wadley, G., Krause, A., Liang, J., Wang, Z., Leong, T.W., 2019. Use of music streaming platforms for emotion regulation by international students, in: Proceedings of the 31st Australian Conference on Human-Computer-Interaction, pp. 337–341.

Wadley, G., Smith, W., Koval, P., Gross, J.J., 2020. Digital emotion regulation. Current Directions in Psychological Science 29, 412–418.

Walther, J.B., 1993. Impression development in computer-mediated interaction. Western Journal of Communication (includes Communication Reports) 57, 381–398.

Wang, Y., Diakopoulos, N., 2021. The role of new york times picks in comment quality and engagement, in: 54th Annual Hawaii International Conference on System Sciences, HICSS 2021, IEEE Computer Society. pp. 2924–2933.

Xenos, A., Pavlopoulos, J., Androutsopoulos, I., Dixon, L., Sorensen, J., Laugier, L., 2021. Toxicity detection can be sensitive to the conversational context. arXiv preprint arXiv:2111.10223 .

Yang, K., Wang, C., Sarsenbayeva, Z., Tag, B., Dingler, T., Wadley, G., Goncalves, J., 2021. Benchmarking commercial emotion detection systems using realistic distortions of facial image datasets. The visual computer 37, 1447–1466.

Yang, S., Xing, L., Li, Y., Chang, Z., 2022. Implicit sentiment analysis based on graph attention neural network. Engineering Reports 4, e12452.

Yang, Y., Liu, K., Li, S., Shu, M., 2020. Social media activities, emotion regulation strategies, and their interactions on people's mental health in covid-19 pandemic. International Journal of Environmental Research and Public Health 17, 8931.

Yu, B., Hu, J., Funk, M., Feijs, L., 2018. Delight: biofeedback through ambient light for stress intervention and relaxation assistance. Personal and Ubiquitous Computing 22, 787–805.

Yue, Z., Zhang, R., Xiao, J., 2022. Passive social media use and psychological well-being during the covid-19 pandemic: The role of social comparison and emotion regulation. Computers in Human Behavior 127, 107050.

Zsido, A.N., Arato, N., Lang, A., Labadi, B., Stecina, D., Bandi, S.A., 2021. The role of maladaptive cognitive emotion regulation strategies and social anxiety in problematic smartphone and social media use. Personality and Individual Differences 173, 110647.